\documentclass[fleqn,12pt,twoside]{article}
\usepackage{espcrc1}

\usepackage{graphicx}
\usepackage[figuresright]{rotating}

\newcommand{\AmS}{{\protect\the\textfont2
  A\kern-.1667em\lower.5ex\hbox{M}\kern-.125emS}}

\title{Energy loss of quarks in deconfined matter at RHIC:
photon-tagged jets, single electron and dilepton spectra
from open charm}

\author{K. Gallmeister\address[FZR]{Forschungszentrum Rossendorf,
        PF 510119, 01314 Dresden, Germany}\thanks{present address:
        Institut f\"ur Theoretische Physik, 
        Universit\"at Giessen, Germany},
        B. K\"ampfer\addressmark\thanks{speaker at QM2002}
        and
        O.P. Pavlenko\address{ITP Kiev, Ukraine}}

\begin{document}

\maketitle

\begin{abstract}
We report a first attempt 
(i) to derive constraints on the energy loss of charm quarks 
in a deconfined medium from the recent PHENIX data 
of single-electron transverse momentum spectra and 
(ii) to estimate the resulting suppression of dileptons from
correlated semi-leptonic decays of open charmed mesons.
The momentum imbalance of photon-tagged light-quark jets
is also considered.
\end{abstract}

\section{INTRODUCTION}

Induced gluon radiation of a fast quark propagating through
a deconfined medium of quarks and gluons causes an energy loss
which should considerably modify various observables in 
relativistic heavy-ion collisions compared to $pp$ collisions.
In such a way the properties of the deconfined medium 
(parton composition and space-time dependent densities etc.)
can be probed. The QCD based theory of the energy loss has
been elaborated by various groups; a few key references are
\cite{GLV,BDMS,Urs,DK} from which other relevant publications
can be traced back. As pointed out in \cite{BDMS_JHEP}
the modified transverse momentum spectrum of final hadrons
at midrapidity appears as a convolution of the energy loss distribution 
and the primary spectrum. To enable a comparison with earlier work
\cite{Gallmeister1,Gallmeister2}, 
we employ here a simplified version
by using a Monte Carlo averaging over traversed path lengths
and by shifting the transverse momentum of a quark 
with energy $E$ and mean free path $\lambda$ before
hadronizing by the mean energy loss according to \cite{BDMS}
\begin{equation}
\Delta E = - \frac{\alpha_s}{3} \zeta
\left\{ 
\begin{array}{rll}
0 & : & L < \lambda \: \mbox{or} \: \mbox{in} \: \mbox{hadron} \: \mbox{matter}\\
\hat q (T_f) L^2 
& : & L < L_c\\ 
\sqrt{\hat q (T_f) E} L 
& : & L > L_c 
\end{array} 
\right.
\end{equation}
where $L$ is the traversed path inside the deconfined medium,
$L_c = \sqrt{E / \hat q( T_f)}$, and $\hat q$ encodes the
transport properties of the medium. 
Remarkable is the apparent
independence of the initial state, i.e. the energy loss depends
on the temperature $T_f$ at which the quark leaves the medium.
As we show below, however, due to life time and geometrical size
effects, a sensitivity on the initial conditions occurs.

\setcounter{footnote}{0}
\section{SINGLE ELECTRONS FROM OPEN CHARM DECAYS}

Using the PYTHIA version 6.206 with charm quark mass parameter
$m_c = 1.5$ GeV, 
intrinsic parton transverse momentum
distribution $\sqrt{\langle k_\perp^2 \rangle} = 2.5$ 
GeV,\footnote{This choice describes the charged hadron and $\pi^0$
spectra measured by PHENIX in peripheral collisions
and is compatible with UA1 data
[K. Gallmeister, C. Greiner, Z. Xu, to be published].}
default $Q$ scale 
and K factor
${\cal K} = 5.7$ 
one gets the charm cross section
$\sigma_{c \bar c}^{NN} = 404$ $\mu$b at $\sqrt{s_{NN}} = 130$ GeV. 
With the hybrid fragmentation scheme (Peterson fragmentation
function with $\epsilon = 0.06$) and
the electron/positron decay channels of charmed hadrons within
PYTHIA the resulting inclusive transverse momentum spectrum
agrees fairly well ($\chi^2_{d.o.f.} = 0.27$ (0.39)) 
with the PHENIX data \cite{PHENIX} when using the
appropriate thickness functions $T_{AA} = 6.2$ (22.6) mb${}^{-1}$
for minimum bias (central) collisions, see Figure 1.\\[-12mm] 
\begin{figure}[htb]
\begin{minipage}[t]{75mm}
\includegraphics[angle=-90,width=75mm]{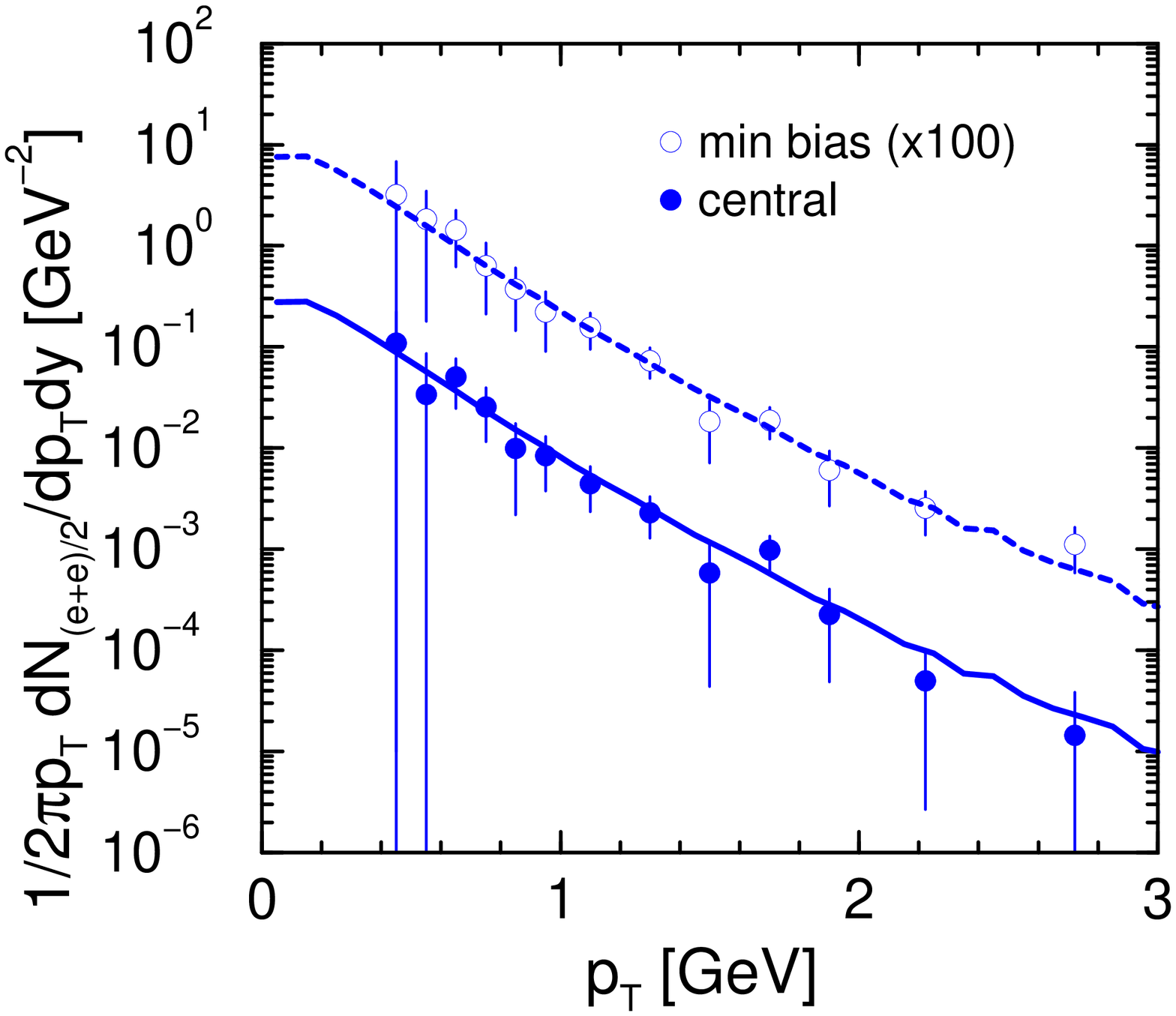}
~\vskip -9mm
\caption{Comparison of our PYTHIA results with the PHENIX data
\protect\cite{PHENIX} (statistical and systematical errors are
quadratically added).}
\label{fig_PHENIX}
\end{minipage}
\hspace{\fill}
\begin{minipage}[t]{75mm}
\includegraphics[angle=-90,width=75mm]{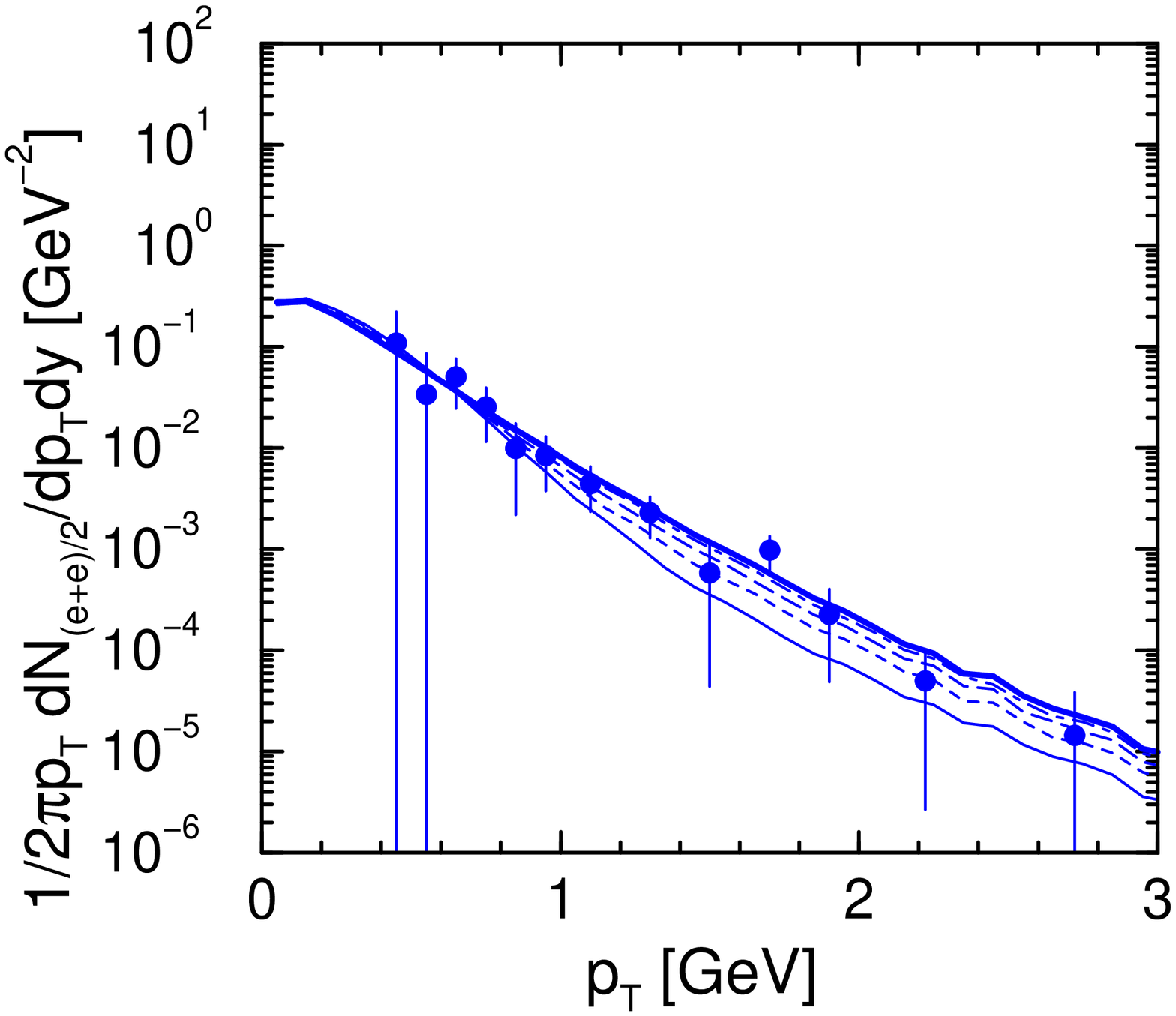}
~\vskip -9mm
\caption{Comparison of various energy loss strengths
$\zeta = 0$, 0.2, 0.5, 1.0, 2.0 (from top to bottom)
with PHENIX data \protect\cite{PHENIX} of central collisions.}
\label{fig_eloss}
\end{minipage}
~\vskip -9mm
\end{figure}

To see which space is left for an energy loss we use the
above described scheme with Bjorken symmetries
(transverse radius $R_A = 7$ fm, no transverse expansion,
full chemical equilibrium, initial time $\tau_i = 0.2$ fm/c) and
initial temperature $T_i = 550$ MeV. The final temperature
$T_f$ depends on the creation point and propagation
direction of the charm quarks; the minimum of $T_f$ is given by
the chiral transition temperature of 170 MeV. We parameterize
different energy loss strengths by $\zeta$. The results of
our Monte Carlo sampling are exhibited in Figure 2. An optimum
description of the data is accomplished by $\zeta = 0.2 \cdots 0.5$,
as quantified by $\chi^2_{d.o.f.} = 0.32 \cdots 0.31$. It turns
out, however, that larger energy losses are also compatible
with data (e.g.,  $\chi^2_{d.o.f.} = 0.43$ (0.76) 
for $\zeta = 1.0$ (2.0)),
as no energy loss does
($\chi^2_{d.o.f.} = 0.39$ for $\zeta = 0$). Insofar, the present data
do not constrain significantly the energy loss of charm quarks.
Our neglect of the dead cone effect \cite{DK} and the use of the
mean energy loss instead of the proper distribution \cite{BDMS_JHEP}
overestimates the theoretical energy loss. We are aware that it 
would be better to compare $pp$ data with central $AA$ data because
the minimum bias data might be contaminated by energy losses.
 
\section{DILEPTON SUPPRESSION}

As pointed out in \cite{Shuryak} and quantified in
\cite{Gallmeister1,Vogt}, energy loss effects
can suppress the dileptons from charm decays. This is a
potentially important effect since these charm contributions
compete with the Drell-Yan yield \cite{Vesa} and hide the
interesting thermal contribution.
Given the above parameterization of the modifications
of inclusive single electrons by energy losses of charm quarks,
we proceed to estimate the possible suppression of dileptons
from correlated semi-leptonic decays of open charm mesons.
Our predictions are displayed in Figure 3 for various values
of the strength parameter $\zeta$. Indeed, the dilepton spectra
are quite sensitive to energy losses, however, assuming a small
loss, as suggested by the above analysis, the corresponding
suppression is small.
This implies that without subtracting the charm component
in dilepton spectra, an identification and quantification
of the thermal yield will hardly be possible.
Otherwise, as shown in \cite{Gale} the thermal dilepton contribution
allows a very concise characterization of the highly excited
strongly interacting matter. 
Therefore, it would be very useful to get experimental 
dilepton spectra with identified charm contribution.\\[-9mm]  
\begin{figure}[htb]
\begin{minipage}[t]{75mm}
\includegraphics[angle=-90,width=75mm]{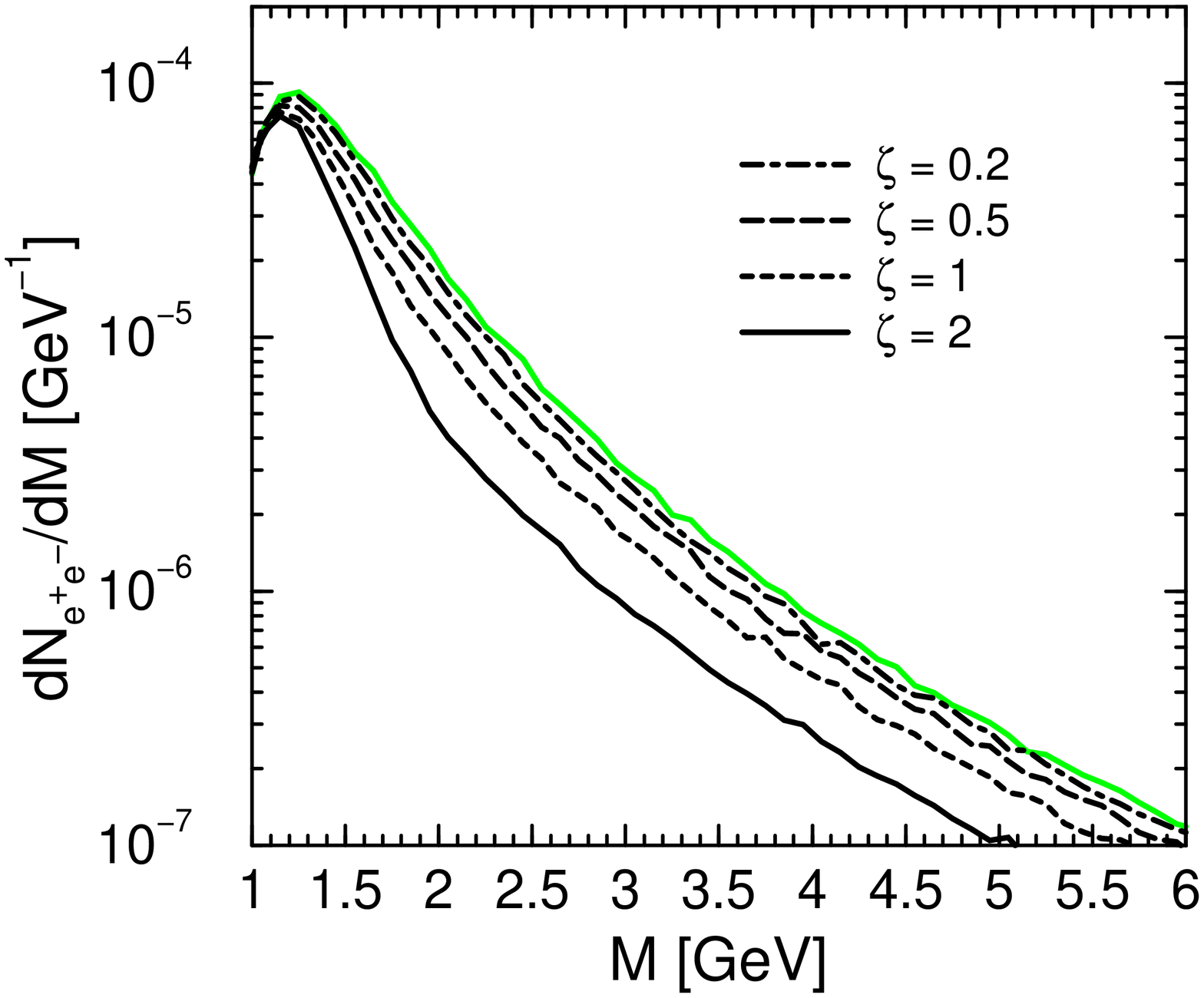}
\label{fig_dileptons_1}
\end{minipage}
\hspace{\fill}
\begin{minipage}[t]{77mm}
\includegraphics[angle=-90,width=75mm]{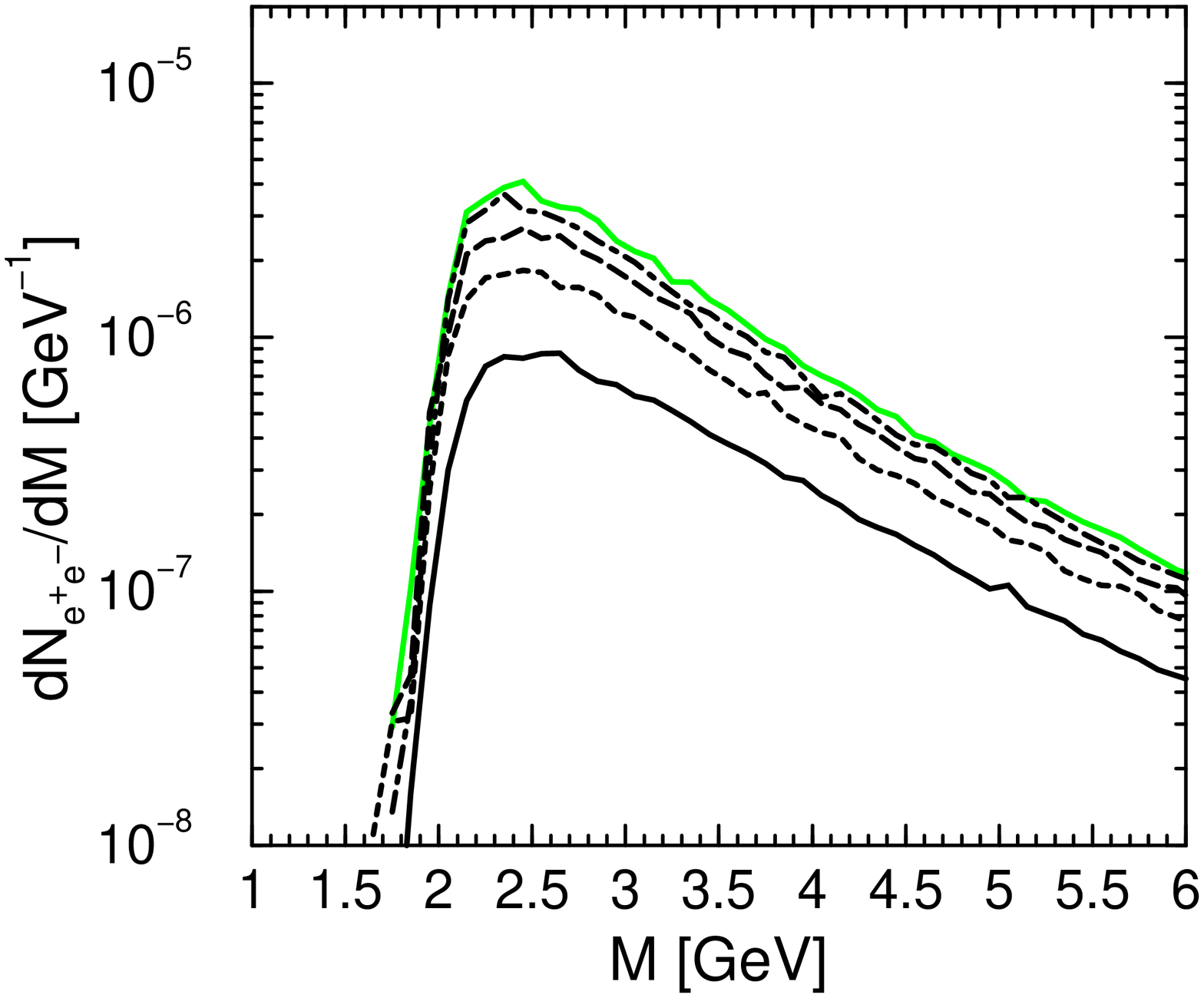}
\label{fig_dileptons_2}
\end{minipage}
~\vskip -9mm
\caption{Predicted dilepton spectra from open charm mesons
for various strength parameters of the energy loss
within the PHENIX acceptance.
$T_{AA} = 31$/mb, $\sqrt{s_{NN}} = 200$ GeV.
Left (right) panel: single-lepton $p_\perp^{\rm min} = 0.5$ (1.0) GeV/c.}
~\vskip -12mm 
\end{figure}

\section{MOMENTUM IMBALANCE OF PHOTON-TAGGED JETS}

Following the suggestion in \cite{Wang} one can try to extract
information on energy losses from photon-tagged jets. After the
hard reaction $g + q \to \gamma + q$ the outgoing photon does
not suffer any noticeable modification by the ambient medium.
Therefore, the momentum imbalance 
$p_\perp^\gamma - \langle p_\perp^q \rangle$ may serve as a 
sensible quantity to characterize the energy loss of the outgoing
quark $q$ which can be identified in a jet by selecting
a sufficiently narrow cone \cite{Gallmeister2}.
Within the above described scheme the resulting momentum imbalance
is depicted in Figure 4. Clearly seen is the dependence on the
initial condition which essentially comes from life time effects. 
For $T_i = 550$ MeV the two-regime behavior from Eq.~(1) is evidenced.
For more details consult \cite{Gallmeister2}.
\newpage
\begin{figure}[htb]
\begin{minipage}[t]{75mm}
\includegraphics[angle=-90,width=75mm]{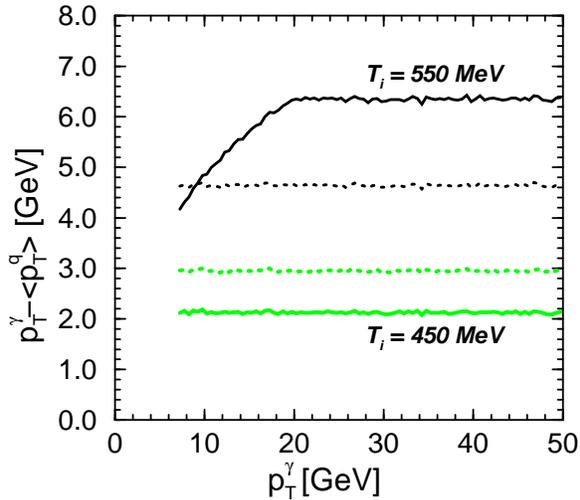}
\label{fig_jet}
\end{minipage}
\hspace{\fill}
\begin{minipage}[t]{75mm}
\caption{Momentum imbalance of photon-tagged jets at midrapidity
for two initial temperatures. $\zeta = 2$ in the energy
loss scheme according to Eq.~(1) is used (solid curves).
The dotted curves  are for a constant
energy loss $dE/dx = - 1$ GeV/fm.}
\end{minipage}
~\vskip -9mm 
\end{figure}

\section{SUMMARY}

The present inclusive single-electron spectra \cite{PHENIX} 
from open charm decays in Au + Au collisions 
at $\sqrt{s_{NN}} = 130$ GeV seem to point
to tiny energy losses. More quantitative conclusions can be drawn
after the release of the data of Au + Au collisions at
$\sqrt{s_{NN}} = 200$ GeV which have better statistics and better
centrality selection \cite{Averbeck}; also the release of the
$pp$ data at the same energy will be very helpful.

Photon-tagged jets are considered useful 
to accomplish the goal of jet tomography of deconfined matter.

Discussions with R. Averbeck, Y. Akiba, and B. Cole
are gratefully acknowledged.

\end{document}